\documentclass[prl,amsmath]{revtex4}
\usepackage{amssymb}
\usepackage{epsfig}

\begin{document}
\draft
\title{Electron-hole imbalance in superconductor-normal metal mesoscopic structures.}
\author{V.R. Kogan$^{1,3},$ V.V. Pavlovskii$^{2}$ and A.F. Volkov$^{1,2}$}
\address{$^{(1)}$Theoretische Physik III,\\
Ruhr-Universit\"{a}t Bochum, D-44780 Bochum, Germany\\
$^{(2)}$Institute of Radioengineering and Electronics of the Russian Academy%
\\
of Sciences, 103907 Moscow, Russia \\
$^{(3)}$L.D. Landau Institute for Theoretical Physics, 117940 Moscow, Russia}

\bigskip
\begin{abstract}
We analysed the electron-hole or, in another words, branch imbalance (BI)
and the related electric potential $V_{imb}$ which may arise in a mesoscopic
superconductor/normal metal (S/N) structure under non-equilibrium conditions
in the presence of a supercurrent. Non-equilibrium conditions can be created
in different ways: a) a quasiparticle current flowing between the N
reservoirs; b) a temperature gradient between the N reservoirs and no
quasiparticle current. It is shown that the voltage $V_{imb}$ oscillates
with the phase difference $\varphi$. In a cross-geometry structure the
voltage $V_{imb}$ arises in the vertical branch and affects the conditions
for a transition into the $\pi -$state.
\end{abstract}
\maketitle
\bigskip\bigskip

A few decades ago a great deal of interest was paid to the study of effects
related to the so called branch imbalance (BI) (see Refs. ~~\cite
{Clarke,TinkClarke} and for example the reviews ~\cite{ArtV,SmPeth}). The BI
implies that populations of the electron-like and hole-like branches of the
excitation spectrum in a superconductor or a normal metal are different. For
example, the BI may arise in a superconductor near the S/N interface if a
current flows through this interface and the temperature is close to $T_{c}.$
The conversion of the quasiparticle current $j_{Q}$ into the condensate
current $j_{S}$ occurs over a rather long length $\lambda _{Q}$ called a BI
relaxation length. Over this length populations of the electron-like and hole-like branches of
the energy spectrum differ from each other. The difference between these populations is characterized by the
distribution function $f_{-}=-(n_{\uparrow }-p_{\downarrow });$ the function
$n_{\uparrow }$ is the distribution function of the electron-like
excitations and $p_{\downarrow }(\epsilon )=1-n_{\downarrow }(-\epsilon )$
is the distribution function of the hole-like excitations. In the
considered case of a spin-independent interaction one has $n_{\uparrow
}=n_{\downarrow }$ and $p_{\uparrow }=p_{\downarrow }$. One can show that the function $f_{-}$ differs
from zero if $divj_{Q}\neq 0.$ The non-zero distribution function $f_{-}$
leads to the appearance of an electric potential $V_{imb}$ in a
superconductor (or in a normal metal) which can be expressed in terms of the
function $f_{-}$ (see below). The BI may also arise in a bulk superconductor. For
example, if longitudinal collective oscillations with a finite wave vector $q
$ are excited in the  superconductor, the BI arises because in this case $%
divj_{Q}=iqj_{Q}\neq 0$ .  When these modes are excited (they are weakly
damped only near $T_{c}$)$,$ the quasiparticle current $j_{Q}$ oscillates in
a counter phase with the condensate current $j_{S}$, so that the total
current remains equal to zero. These oscillations have been observed
experimentally by Carlson and Goldman (Carlson-Goldman mode) ~\cite{CG} and
have been explained theoretically in Refs.~\cite{ArtVolk,SS1}. Another
example of a system, in which the BI arises, is a uniform superconducting
film in the presence of a temperature gradient $\nabla T$ and a condensate
flow. It was established experimentally ~\cite{ClarkeLin} and theoretically ~
\cite{SS2,Peth,Shel} that in this case the BI has a magnitude which is
proportional to $v_{s}\nabla T$, where $v_{s}$ is the condensate velocity.

Recently there has been growing interest in the study of transport
properties of S/N mesoscopic structures. Several interesting, phase-coherent
effects have been observed in these structures. Among them one can mention
the change of sign of the Josephson critical current $I_{c}$ in a
four-terminal S/N/S mesoscopic structure. If an additional dissipative
current (or an applied voltage) between the N reservoirs in a S/N/S
structure of a cross geometry exceeds a certain value,\ the current $I_{c}$
changes sign ~\cite{Wees2} and the Josephson junction is converted into the $%
\pi $-state( a theory for this effect was developed in Refs.~\cite
{Volk,Yip,Wilh}).

\begin{figure}[t]
\centerline{\psfig{figure=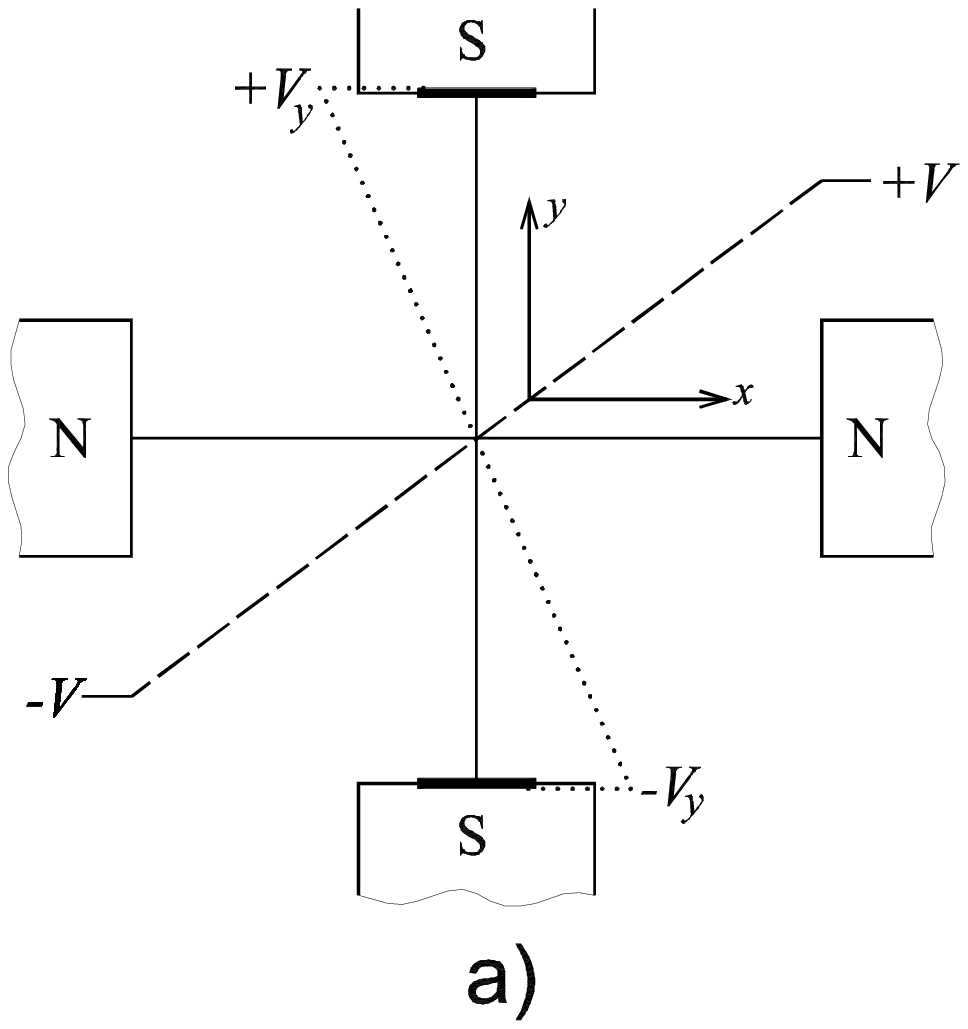,width=8cm,height=6cm}
            \psfig{figure=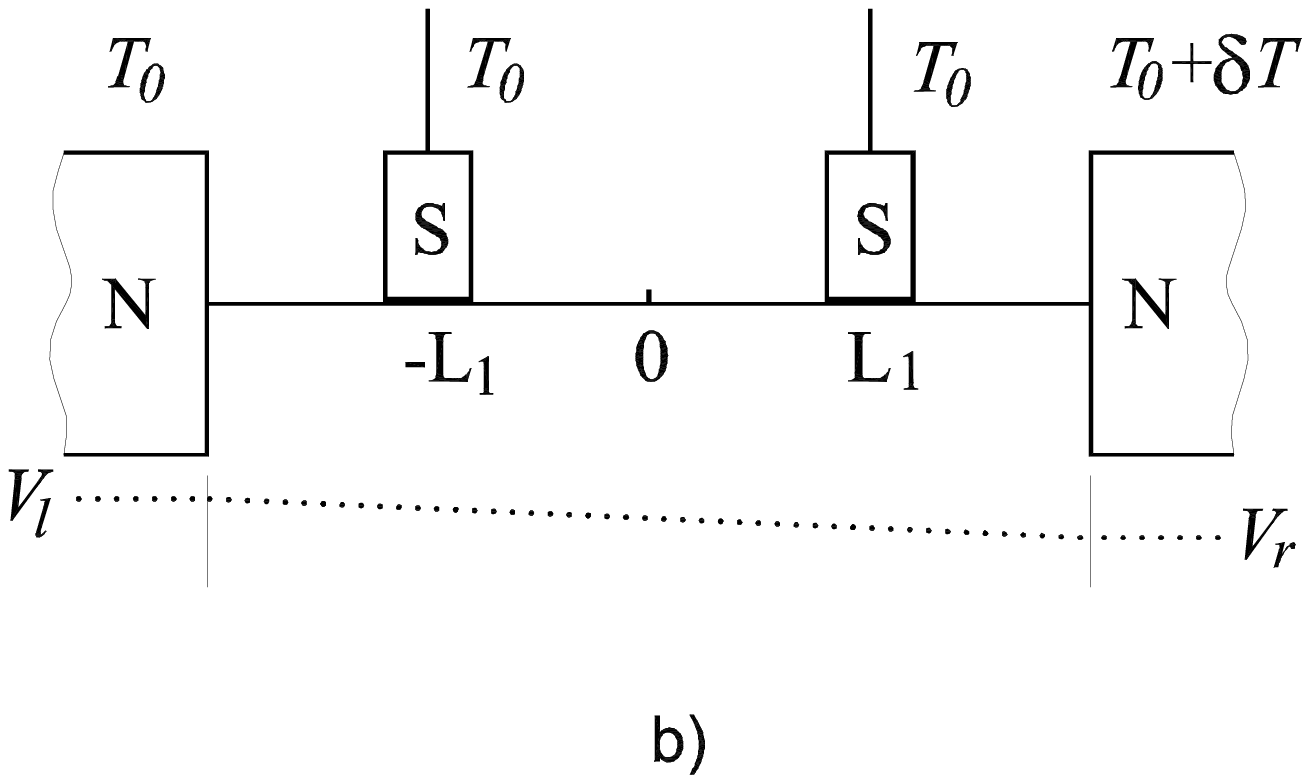,width=8cm,height=6cm}}
\caption{ Schematic view of the structures under consideration. Barriers or
bad S/N interfaces are represented by thick lines. The dashed and dotted
lines show the spatial distribution of the applied voltage (between the N
reservoirs,fig.~\ref{fig1}a) and the voltage caused by the charge imbalance.
The electric potential in the superconductors is assumed to be zero. }
\label{fig1}
\end{figure}

In this Letter, we consider S/N mesoscopic structures under non-equilibrium
conditions and study the BI which can arise in the N conductor if there is a
finite phase difference $\varphi $ between the superconductors in these
structures. We briefly analyse effects arising due to the BI and discuss how
some of them can be observed. To our knowledge these effects are missed in
most papers on transport properties of mesoscopic S/N structures. In some
cases the physics of the BI in mesoscopic S/N structures resembles that of
the BI in a uniform superconductor under nonequilibrium conditions. However
even in these cases the BI has its own specific characteristics. For
example, the voltage $V_{imb}$, which is associated with the BI, is an
oscillating function of the phase difference $\varphi$ between the
superconductors and it may appear even in the absence of a temperature
gradient. Consider for example the structure in fig.~\ref{fig1}a.
We will show that the voltage $V_{imb}(y)$ (or the electric field) arises in
the vertical wire regardless of how the system is brought out from the
equilibrium state. One can apply a temperature gradient between the N
reservoirs disconnected from the external circuit or one can pass a finite
current between the N reservoirs as it was done in the experiment ~\cite
{Wees2}. When considering S/N/S structures, it is convenient to introduce
two types of the distribution functions: $f_{\pm }$ as it was done in Ref. ~
\cite{LO}. The function $f_{+}$ determines the order
parameter in the superconductors and, for example, the condensate current in
S/N/S structures. This function is related to the distribution functions of
electrons $n_{\uparrow }=n_{\downarrow }$ and holes $p_{\uparrow
}=p_{\downarrow }$: $f_{+}=1-(n_{\uparrow }+p_{\downarrow })$%
\textbf{. }The Josephson current in the N wire can be written in the form
\begin{equation}
j_{s}=(\sigma /2)W\int d\epsilon f_{+}J_{s}  \label{eq1}
\end{equation}
where $W$ is the cross-section area of the y-wire, the distribution function
$f_{+}$, generally speaking, deviates from its equilibrium form $%
f_{eq}=\tanh \epsilon \beta $ and should be determined from the kinetic
equation, here $\beta =1/2T.$ The function $J_{s}$ is ''a partial condensate
current'' and is expressed in terms of the condensate Green's functions $%
\widehat{F}^{R(A)}$: $J_{s}=(1/4)Tr\{\sigma _{z}(\widehat{F}^{R}\partial _{y}%
\widehat{F}^{R}- \widehat{F}^{A}\partial _{y}\widehat{F} ^{A})\}$. This
"current" does not depend on $y$. On the other hand, as follows from the
conservation of the current, $j_{s}$ can be written as a current through the
S/N interface. This means in particular that "the current" $J_{s}$ can be
represented in the form $J_{s}=(r_{s}/L_{y})Tr\{\widehat{ \sigma} _{z}(%
\widehat{F}^{R}\widehat{F}_{s}^{R}-\widehat{F}^{A}\widehat{F} _{s}^{A})\}$,
where the functions $\widehat{F}^{R(A)}$ must be taken at the S/N
interfaces; $r_{s}=R_{y}/R_{s}$, $R_{y}=\varrho L_{y}$ is the resistance of
the $y$-branch (per unit area) and $R_{s}$ is the the resistance of the S/N
interface (per unit area). Under nonequlibrium conditions, which may have a
different origin (non-equal temperatures of the S and N reservoirs, an
additional current between the N reservoirs etc), the current given by eq.~(%
\ref{eq1}) is not equal to the Josephson current through the S/N interface
because the last one depends not only on the value of $f_{+}$ at the
interface but also on the distribution function $f_{S+}$ in the S reservoirs
which is assumed to be equilibrium ($f_{S+}=f_{eq}$). This means that an
electric field $E=-\partial _{y}V_{imb}$ arises in the $y$-branch even if
there are no voltage differences neither between N reservoirs nor between
the S reservoirs. The electric field drives the quasiparticle current which
compensates the above mentioned difference between the condensate currents.
The appearance of the electric field and the quasiparticle current means
that strictly speaking eq.~(\ref{eq1}) does not describe the maximum current
in the $y$-branch in the absence of the voltage difference between the S
reservoirs (in most papers on this subject the maximum current was found
from eq.~(\ref{eq1})). In order to find the critical current in a
non-equilibrium situation, one has to use a more general formula. However at
low temperatures eq.~(\ref{eq1}) determines the critical current $I_{c}$
with a good (exponential) accuracy.

In the present paper we consider S/N mesoscopic structures of two
configurations (see fig.~\ref{fig1}). In simple limiting cases we find the
distribution functions and analyse the BI arising under non-equilibrium
conditions. The distribution functions $f_{\pm }$ obey the kinetic equations
~\cite{LO} which for the structure shown in fig.~\ref{fig1}b can be written
in the form (see for example ~\cite{VSev})
\begin{equation}
L{\ \partial }_{x}[{\ M}_{\pm }{\ \partial }_{x}f_{\pm }(x){\ +J}_{S}f_{\mp
}(x)\pm {\ J}_{an}{\ \partial }_{x}f_{\mp }(x)]{\ =r}_{S}{\ [}\stackrel{}{%
A_{\pm }}{\ \delta (x-L}_{1}{\ )+}\stackrel{\_}{A_{\pm }}{\ \delta (x+L}_{1}{%
\ )].}  \label{eq2}
\end{equation}
where all the coefficients are expressed \ in terms of the retarded
(advanced) Green's functions: $\widehat{G}^{R(A)}=G^{R(A)}\hat{\sigma}_{z}+%
\widehat{F}^{R(A)};$ $\ M_{\pm }=(1-G^{R}G^{A}\mp (\widehat{F}^{R}\widehat{F}%
^{A})_{1})/2;$ $J_{an}=(\widehat{F}^{R}\widehat{F}^{A})_{z}/2,\ J_{s}=(1/2)(%
\widehat{F}^{R}\partial _{x}\widehat{F}^{R}-\widehat{F}^{A}\partial _{x}%
\widehat{F}^{A})_{z},$\ $A_{\pm }=(\nu \nu _{S}+g_{1\mp })(f_{\pm }-f_{S\pm
})-(g_{z\pm }f_{S\mp }+g_{z\mp }f_{\mp });$ $g_{1\pm }=(1/4)[(\widehat{F}%
^{R}\pm \widehat{F}^{A})(\widehat{F}_{S}^{R}\pm \widehat{F}_{S}^{A})]_{1};$ $%
g_{z\pm }=(1/4)[(\widehat{F}^{R}\mp \widehat{F}^{A})(\widehat{F}_{S}^{R}\pm
\widehat{F}_{S}^{R})]_{z}.$ The parameter $r_{S}=R_{1}/R_{S}$ is the ratio
of the resistance of the N wire $R_{1}=\rho L_{1}$ and S/N interface
resistance $R_{S}$; the functions $\stackrel{\_}{A_{-}}$ and $\stackrel{\_}{A%
}_{+}$ coincide with $\stackrel{}{A_{-}},\stackrel{}{A}_{+}$ if we make a
substitution $\varphi \rightarrow -\varphi $. We introduced above the
following notations $(\widehat{F}^{R}\widehat{F}^{A})_{1}=Tr(\widehat{F}^{R}%
\widehat{F}^{A})/2,$ $(\widehat{F}^{R}\widehat{F}^{A})_{z}=Tr(\hat{\sigma}%
_{z}\widehat{F}^{R}\widehat{F}^{A})/2$ etc.; $\nu ,$ $\nu _{S}$ are the
density-of states in the N film at $x=L_{1}$ and in the superconductors. The
functions $f_{S\pm }$ are the distribution functions in the superconductors
which are assumed to have the equilibrium forms. This means that $%
f_{S+}\equiv f_{eq}=\tanh (\epsilon \beta _{o})$ and $f_{S-}=0,$ because we
set the potential of the superconductors equal to zero. We neglect branch
imbalance in the superconductors assuming that the distribution functions $%
f_{\pm }$ recover quickly their equilibrium forms in S due to a big size of
the superconductors in comparision with the size of the S/N interface. In
the case of the structure in fig.~\ref{fig1}a the left-hand side of eq.~(\ref
{eq2}) should be written down for each branch of the structure and be set
equal to zero. At the S/N and N/N' interfaces we use the boundary conditions
which are given by the right-hand side of eq.~(\ref{eq2}) (at the N/N'
interfaces the index $S$ should be replaced by the index $N^{\prime }$ and
all the condensate Green's functions in N' should be set equal to zero).
Consider the structure shown in fig.~\ref{fig1}a. Eqs.~(\ref{eq1}) can be
integrated once in each branch, and, for example, in the $y$-branch we
obtain
\begin{equation}
{\ M}_{\pm }{\ \partial }_{y}f_{\pm }(y){\ +J}_{S}f_{\mp }(y)\pm {\ J}_{an}{%
\ \partial }_{y}f_{\mp }(y){\ =J}_{y\gtrless }{.}  \label{eq3}
\end{equation}
where $J_{y\gtrless }$ are the total ''partial currents'' in the upper ($%
y>0) $ and lower ($y<0)$ \ parts of the $y$-branch of the N wire. The
current $J_{y\gtrless }$ is a ''vector'' with the components ($J_{y+},J_{y-}$
). At the crossing point the current conservation law takes place $%
J_{y>}+J_{x>}=J_{y<}+J_{x<}.$ At the S/N interface the current $J_{y>}$ is
related to the Green's functions at $y=L_{y}$
\begin{equation}
{\ J}_{y>}=(r_{S}/L_{y}){\ A.}  \label{eq5}
\end{equation}

In order to obtain the current through the y-wire, the "current" $J_{y>}$
should be substituted into the integrand of eq.~(\ref{eq1} instead of \ $%
f_{+}J_{s}$. We present the solutions  for the distribution functions
assuming first the weak proximity effect. This means that the amplitude of
the condensate functions in the N wire should be small:\ $|F^{R(A)}|<<1.$ In
this case $F^{R(A)}$ can be easily found from the linearized Usadel equation
(the solution for the structure shown in fig.~\ref{fig1}b is presented in
ref.~\cite{ThermVS}). As follows from the form of the functions $F^{R(A)}$,
they are small if the condition $\varepsilon >>\varepsilon _{y}r_{S}$ is
satisfied, where the characteristic energy $\varepsilon $ is equal to the
Thouless energy $\varepsilon _{y}=D/L_{y}^{2}$ ( in the case of the geometry
in fig.~\ref{fig1}b the Thouless energy is $\varepsilon _{1}=D/L_{1}^{2})$
or to the temperature $T.$ In the case of the weak proximity effect, a
solution for the kinetic equations ~(\ref{eq1}) also can be easily found
with the help of expansion in the parameter $r_{S}$ (for the case $r_{S}>1$
eq.~(\ref{eq2}) was solved numerically in Ref.~\cite{Yip}). We consider two types of the nonequalibrium situation: a) the
temperatures of all the reservoirs are the same, but the electric potentials
at the left and right N reservoirs are $\pm V$ (a current flows between
these reservoirs); b) no current between the N reservoirs, but the
temperatures of the left and right N reservoirs are different, so
that the distribution function in the right (left) reservoirs $F_{r,l}$%
is equal to: $F_{r,l}=\tanh (\epsilon \beta _{r,l})$. In the main approximation we find
\begin{eqnarray}
a)f_{+}(x) &=&F_{V+}; f_{+}(y)=F_{V+}  \nonumber \\
b)f_{+}(x)&=&(F_{l}+F_{r})/2+(x/L_{x})(F_{r}-F_{l})/2;
f_{+}(y)=(F_{l}+F_{r})/2  \label{eq6}
\end{eqnarray}

and
\begin{eqnarray}
a)f_{-}(x) &=&(x/L_{x})F_{V-}; f_{-}(y)=r_{S}(y/L_y)g_{z-}(f_{eq}-F_{V+});
\nonumber \\
b) f_{-}(x) &=&0; f_{-}(y)=r_{S}(y/L_y)g_{z-}[f_{eq}-(F_r+F_l)/2]
\label{eq7}
\end{eqnarray}

\begin{figure}[t]
\centerline{\psfig{figure=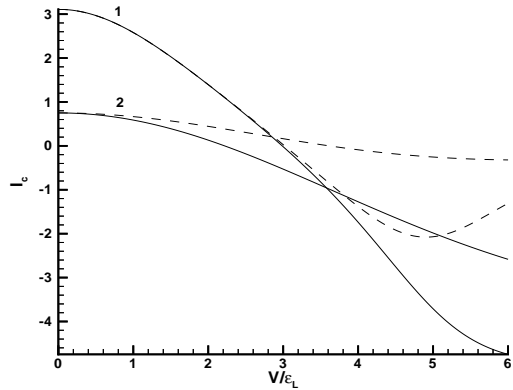,width=8cm,height=6cm}}
\caption{ The dependence of the critical current $I_{c}$ (in arbitrary
units) on the applied voltage $V$ for two temperatures: $T=0.5\protect%
\epsilon _{L}$ and $T=2\protect\epsilon _{L}$, where $\protect\epsilon %
_{L}=D/L^{2},L_{y}=0.3L$. The value $\protect\epsilon _{L}$ is chosen in
such a way that $\Delta(0)=5\protect\epsilon_{L}$. The solid lines are
calculated from the correct eq.~(\ref{eq5}) and the dashed lines are
calculated from the approximate, widely used eq.~(\ref{eq1}). The difference
between these curves determines the voltage $V_{y}$ (divided by the S/N
interface resistance) at the S/N interface. }
\label{fig2}
\end{figure}

Knowing the distribution function $f_{-}$, we can calculate the electric
potential in the $y$-branch $V$ with the help of the formula ~\cite{LO,ArtV}

\begin{equation}
eV(y)=(1/8)Tr\int d\epsilon \widehat{G}(\epsilon ,y)=(1/2)\int d\epsilon
\upsilon (\epsilon )f_{-}  \label{eq8}
\end{equation}
where $\widehat{G}=\widehat{G}^{R}\widehat{f}-\widehat{f}\widehat{G}^{A}$ is
the Keldysh component of the matrix Green's function, $\widehat{f}=\widehat{%
1 }f_{+}+\widehat{\sigma }_{z}f_{-}$ is the matrix distribution function, $%
\upsilon (\epsilon )=(1/4)Tr(\widehat{\sigma }_{z}(\widehat{G}^{R}-\widehat{G%
}^{A}))$ is the density-of-states in the N wire. We easily find from eqs.~(%
\ref{eq7}-\ref{eq8}) that the potential $V(y)$ is an odd function of $y$ and
the electric field $E(y)$ is an even function of $y$. We also see that the
electric field arises only if the phase difference between the
superconductors is not zero; otherwise the function $g_{z-}$ is zero. The
field $E$ or the potential $V(y)$ oscillate with increasing phase difference
$\varphi$. As is seen from eqs.~(\ref{eq7}-\ref{eq8}), the electric field
arises regardless of the origin of the non-equilibrium state: the function $%
f_{+}$ may deviate from the equilibrium function $f_{eq}$ if a finite
current flows between the N reservoirs or if the temperature of the N
reservoirs differs from the temperature of the S reservoirs. In the
considered case of a weak proximity effect the conductance $G$ between the N
reservoirs decreases with increasing the phase difference $\varphi $ (at
small $\varphi $). We also have calculated the critical (maximum) current
using the correct expression eq.~(\ref{eq5}) and the approximate one eq.~(\ref{eq1})
(see fig.~\ref{fig2}).

We see that the difference between two curves is significant if the
temperature is not low. The difference between the critical currents is
determined by the quasiparticle current $j_{Q}=\sigma E$. This follows
directly from eq.~(\ref{eq3}) if one takes into account that in the main
approximation $M_{-}\cong 1$ and the last term on the right-hand side can be
neglected. It is worth noting that $f_{-}\neq 0$ only in
the normal wire. In the superconductors the function $f_{-}$ and
therefore the voltage $V$ are assumed to be zero, so that at the
S/N interface there is a voltage drop from a finite value of $V$
in the N wire to zero in the superconductor.

We also considered another limiting case when the proximity effect is not
weak, that is, the condensate function in the N wire $\widehat{|F}|$ is not
small. The obtained results qualitatively are similar to those which have
been established for the weak proximity effect. The only difference is that
the conductance in this case increases with increasing phase difference $%
\varphi$. Different behaviour of the conductance $G$  as a function of $%
\varphi$ was studied in Refs.~\cite{VZ} where a transition from a decreasing
to increasing dependence (at small $\varphi$) of $G(\varphi)$ was obtained
by varying the applied voltage. In a recent paper ~\cite{Belz} a similar
transition (obtained by varying the temperature) was studied in detail both
experimentally and theoretically. Contrary to Refs. ~\cite{VZ} it was
assumed in Ref.~\cite{Belz} that the S/N interface is perfectly transparent.
A good agreement between theoretical results and experimental data was
obtained.

\begin{figure}[t]
\centerline{\psfig{figure=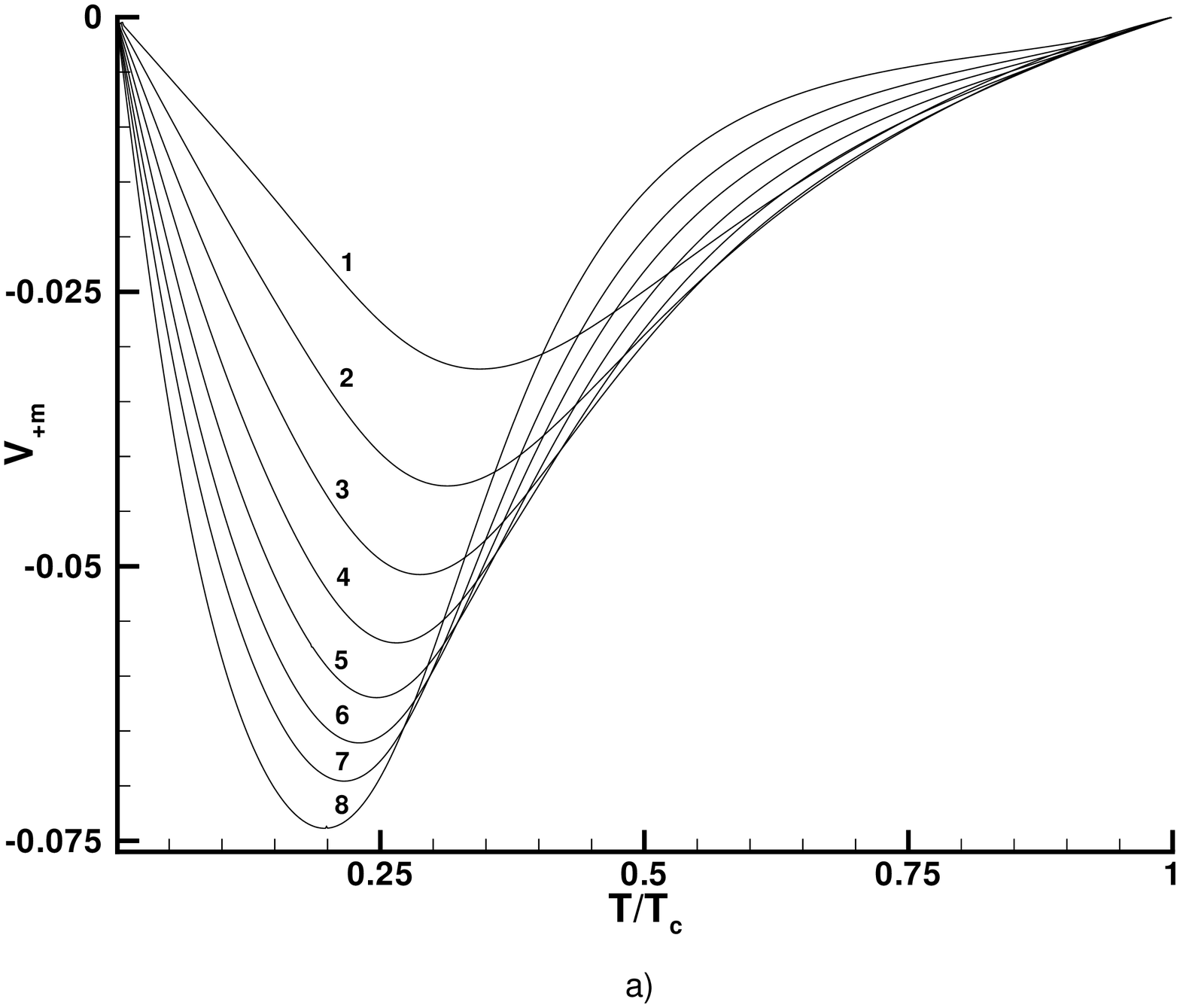,width=8cm,height=6cm}
            \psfig{figure=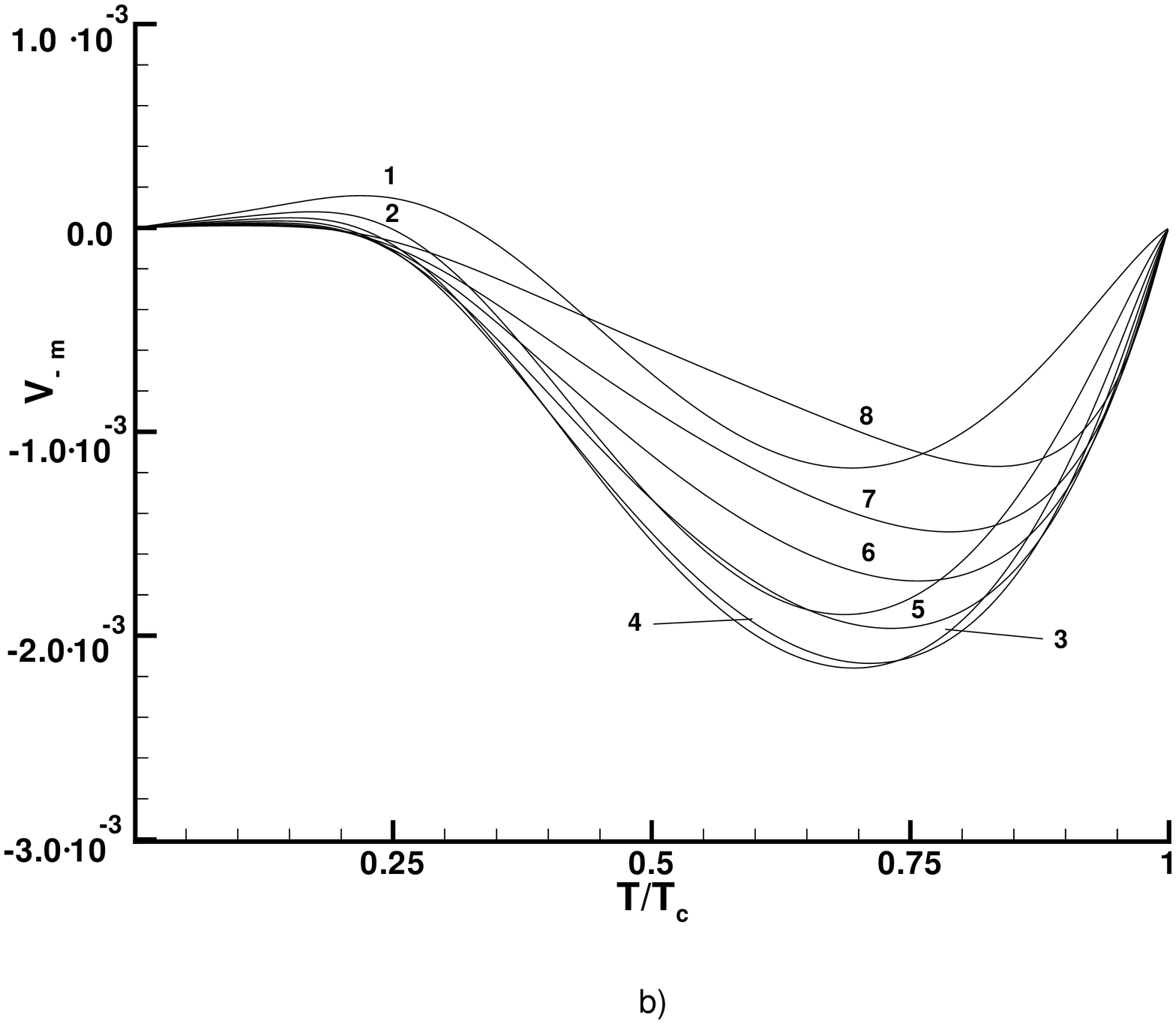,width=8cm,height=6cm}}
\caption{ The temperature dependence of the voltages $V_{+}=(V_{r}+V_{l})/2$
and $V_{-}=(V_{r}-V_{l})$ caused by an extraordinary, phase-coherent
thermoemf for different values of the parameter $\Delta (0)/\protect\epsilon
_{L}$: $\Delta (0)/\protect\epsilon _{L}=1.0;1.4;1.8;2.2;2.6;3.0;3.4$ and $%
4.0$ for the curves $1-8$, respectively. }
\label{fig3}
\end{figure}

At last we consider a specific thermoelectric effect arising in the
structure in fig.~\ref{fig1}b. This effect was measured recently in
mesoscopic S/N structures ~\cite{ThermPetr,ThermChandra}. As established in ~
\cite{ThermVS}, if temperatures of the normal reservoirs are different, a
voltage arises between these and superconducting reservoirs (the S
reservoirs have the same electric potential because they are connected with
a superconducting loop). The origin of this voltage which can be called
thermoemf is completely different from the ordinary thermoemf in S/N/S
junctions studied in Refs. ~\cite{Gal}. In the last case the thermoemf
appears due to the ordinary thermoelectric component of the quasiparticle
current which is neglected here. In the limit of the weak proximity effect
we have calculated the distribution functions $f_{\pm }$ and the voltages $%
V_{l,r}$ caused by the temperature difference between the N reservoirs; here
$V_{l,r}$ are the electric potentials at the left and right N reservoirs,
respectively (the electric potential at the S reservoirs is set to zero). We
assumed that the temperature of the left N reservoir $T_{o}$ coincides with
the temperature of the S reservoirs and the temperature of the right N
reservoir\ $T$ is elevated: $T=T_{o}+$ $\delta T$. The distribution
functions can be easily found using an expansion in the parameter $r_{S}$.
For the voltages $V_{\pm }=(V_{r}\pm V_{l})/2$, we obtain from eq.~(\ref{eq8}%
)
\begin{eqnarray}
eV_{+}=-\delta T(L_{1}/L)\int d\epsilon (\epsilon \beta )
g_{z+}(\epsilon,L_{1})f_{eq}^{\prime }/\int d\epsilon g_{1+}(\epsilon
,L_{1}) f_{eq}^{\prime};  \nonumber \\
eV_{-}=r_{S}\delta T(L_{1}/2L)\int d\epsilon (\epsilon \beta )
g_{z-}(\epsilon ,L_{1})f_{eq}^{\prime }  \label{eq11}
\end{eqnarray}
where $f_{eq}^{\prime }=\cosh ^{-2}(\epsilon \beta ).$ The expression for $%
V_{+}$ was obtained in Ref.~\cite{ThermVS}. Here we also obtained the
voltage difference between the N reservoirs $V_{-}$. One can see that this
voltage contains the small parameter $r_{S}$ in comparision to the voltage $%
V_{+}.$ Both voltages are proportinal to $\sin \varphi $, that is, they
oscillate with increasing phase difference. The temperature dependence of
the amplitudes of $V_{\pm }$ are plotted in fig.~\ref{fig3}.

We see that the maximum of the amplitude of $V_{+}$ is located at a lower
temperature than the maximum of $V_{-}$. In principle this behaviour may
lead to a nonmonotonic behaviour of the voltages $V_{r}$ or $V_{l}$ as a
function of temperature. \ The change of the phase of the voltage
oscillations with increasing phase $\varphi$ may have the same origin as
that in the case of the conductance ~\cite{Belz}, i.e. the change of the
dependence $\varphi (H)$,where $H$ is the external magnetic field$.$ Our
results can not be compared quantitatively with the recent experimental data
~\cite{ThermPetr} because the experimental structure corresponds to the case
$r_{S}\gtrapprox 1$. We will analyse this more complicated case in a
separate paper.

In conclusion, we have analyzed the branch imbalance effects in S/N
mesoscopic structures. We have shown that in the structure in fig.~\ref{fig1}%
a a voltage $V_{y}$ related to the BI is set up if a current is driven
through the x-axis or a temperature gradient exists between the N
reservoirs. The voltage $V_{y}$ is proportional to $\sin \varphi $, i.e. it
oscillates with increasing phase difference $\varphi .$ A similar voltage $%
V_{imb}$ arises in the vertical wire if a temperature gradient $\delta T$
exists between the N reservoirs. We also studied an unusual thermoelectric
effect in the structure shown in fig.~\ref{fig1}b. In this case voltages $%
V_{r,l}$ arise in the right and left N reservoirs (the electric potential at
the superconductors is assumed to be zero) if there is a temperature
difference between the normal resevoirs. These voltages are proportional to $%
\delta T\sin \varphi$ and they are not related to the ordinary thermoemf
because we have ignored the small thermoelectric component in the
quasiparticle current.

\acknowledgments We would like to thank SFB 491 and GRK 384 for financial
support.

\end{document}